\documentclass[12pt]{article}
\usepackage{amssymb}

\newcommand{\be}{\begin{equation}}
\newcommand{\ee}{\end{equation}}
\newcommand{\bea}{\begin{eqnarray}}
\newcommand{\eea}{\end{eqnarray}}
\newcommand{\barray}{\begin{array}}
\newcommand{\earray}{\end{array}}

\newcommand{\nn}{\nonumber}
\newcommand{\bitem}{\begin{itemize}}
\newcommand{\eitem}{\end{itemize}}
\newtheorem{teo}{Theorem}[section]
\newcommand{\bt}{\begin{teo}}
\newcommand{\et}{\end{teo}}
\newtheorem{Def}{Definition}[section]
\newcommand{\bd}{\begin{Def}}
\newcommand{\ed}{\end{Def}}
\newtheorem{lem}{Lemma}[section]
\newcommand{\bl}{\begin{lem}}
\newcommand{\el}{\end{lem}}
\newtheorem{prop}{Proposition}[section]
\newcommand{\bp}{\begin{prop}}
\newcommand{\ep}{\end{prop}}
\newtheorem{cor}{Corollary}[section]
\newcommand{\bc}{\begin{cor}}
\newcommand{\ec}{\end{cor}}
\newtheorem{ex}{Example}[section]
\newcommand{\bex}{\begin{ex}}
\newcommand{\eex}{\end{ex}}
\newtheorem{rem}{Remark}[section]
\newcommand{\br}{\begin{rem}}
\newcommand{\er}{\end{rem}}

\begin{document}

\begin{center}
{\Large \textbf{Consistency on cubic lattices \\ for determinants of
arbitrary orders}}
\end{center}

\medskip

\begin{center}
{\large \bf {O. I. Mokhov}}
\end{center}

\bigskip

\begin{center}
\bf {Abstract}
\end{center}

We consider a special class of two-dimensional discrete equations
defined by relations on elementary $N \times N$ squares, $N
> 2$, of the square lattice $\mathbb{Z}^2$, and propose a new type of
consistency conditions on cubic lattices for such discrete equations
that is connected to bending elementary $N \times N$ squares, $N
> 2$, in the cubic lattice $\mathbb{Z}^3$. For an
arbitrary $N$ we prove such consistency on cubic lattices for
two-dimensional discrete equations defined by the condition that the
determinants of values of the field at the points of the square
lattice $\mathbb{Z}^2$ that are contained in elementary $N \times N$
squares vanish.

\bigskip

\begin{center}
{\large \bf {Introduction}}
\end{center}

\medskip

In this paper we consider a special class of two-dimensional
discrete equations defined by relations on elementary $N \times N$
squares of the square lattice $\mathbb{Z}^2$, $N > 2$, and new,
modified, consistency conditions on cubic lattices for such
equations (these consistency conditions were proposed recently by
the present author in [1]). We also give proofs of the theorems
announced in [1] on the consistency on cubic lattices for
determinants of arbitrary orders.

We consider the square lattice $\mathbb{Z}^2$ consisting of points
with integer coordinates in $\mathbb{R}^2 = \{ (x_1, x_2)|\ x_k \in
\mathbb{R},\ k =1, 2 \}$ and complex (or real) scalar fields $u$ on
the lattice $\mathbb{Z}^2$, $u : \mathbb{Z}^2 \rightarrow
\mathbb{C}$, defined by their values $u_{i_1 i_2}$, $u_{i_1 i_2} \in
\mathbb{C}$, at each lattice point with coordinates $(i_1, i_2)$,
$i_k \in \mathbb{Z}$, $k = 1, 2$.

We consider a class of two-dimensional discrete equations on the
lattice $\mathbb{Z}^2$ for the field $u$ that are given by functions
$Q (x_1, x_2, x_3, x_4)$ of four variables with the help of the
relations \be Q (u_{i j}, u_{i + 1, j}, u_{i, j + 1}, u_{i + 1, j +
1}) = 0, \ \ \ \ i, j \in \mathbb{Z}, \label{1}\ee so that in each
{\it elementary $2 \times 2$ square of the lattice $\mathbb{Z}^2$},
i.e., in each set of lattice points with coordinates of the form
$\{(i, j), (i + 1, j), (i, j + 1), (i + 1, j + 1)\}$, $i, j \in
\mathbb{Z}$, the value of the field $u$ at one of the vertices of
the square is determined by the values of the field at the other
three vertices.

In this case the scalar field $u$ on the lattice $\mathbb{Z}^2$ is
completely determined by fixing initial data, for example, on the
coordinate axes of the lattice, $u_{i \, 0}$ and $u_{ 0 j}$, $i, j
\in \mathbb{Z}$.

Initial data for discrete equations of the form (\ref{1}) can also
be specified, for example, at the points of the lattice
$\mathbb{Z}^2$ that form a "staircase" on the lattice, $u_{k, - k +
s}$, $u_{k + 1, - k + s},$ $k \in \mathbb{Z}$, where $s$ is an
arbitrary fixed integer, $s \in \mathbb{Z}$.

The staircase of points of the lattice $\mathbb{Z}^2$ for initial
data of discrete equations of the form (\ref{1}) can also have the
form $\{(k + s, k), (k + s, k + 1), k \in \mathbb{Z}\}$, where $s$
is an arbitrary fixed integer, $s \in \mathbb{Z}$. Moreover, the
staircase of points of the lattice $\mathbb{Z}^2$ for initial data
of discrete equations of the form (\ref{1}) can also have steps of
arbitrary sizes of the form $\{(k, s_k + j), k \in \mathbb{Z}, 0
\leq j \leq s_{k - 1} - s_k\}$, where $s_k$ is an arbitrary fixed
nonincreasing sequence of integers, $k \in \mathbb{Z}$, $s_k \in
\mathbb{Z}$, $ \ldots \geq s_{k - 1} \geq s_k \geq s_{k + 1} \geq
\ldots$, which is unbounded in both directions, or of the form
$\{(k, s_k + j), k \in \mathbb{Z}, 0 \leq j \leq s_{k + 1} - s_k\}$,
where $s_k$ is an arbitrary fixed nondecreasing sequence of
integers, $k \in \mathbb{Z}$, $s_k \in \mathbb{Z}$, $ \ldots \leq
s_{k - 1} \leq s_k \leq s_{k + 1} \leq \ldots$, which is unbounded
in both directions. It is not difficult to describe all structures
of points of the lattice $\mathbb{Z}^2$ such that fixing initial
data for discrete equations of the form (\ref{1}) at these points
completely determines a scalar field $u$ on the lattice
$\mathbb{Z}^2$.

Here, we do not discuss conditions on the initial data $u_{i j}$
themselves that must correctly and completely determine a scalar
field $u$ on the lattice $\mathbb{Z}^2$ for concrete discrete
equations of the form (\ref{1}), and also we do not discuss
conditions on the functions $Q (x_1, x_2, x_3, x_4)$ for which
relations (\ref{1}) correctly determine a two-dimensional discrete
equation on the lattice $\mathbb{Z}^2$ for the field $u$.

{\it Integrable nonlinear discrete equations} are of particular
importance. In [2]--[4] a special class of integrable discrete
equations of the form (\ref{1}) is singled out by the remarkable and
very natural condition of {\it consistency on cubic lattices} (see
also [5]--[13]).

We consider the cubic lattice $\mathbb{Z}^3$ consisting of points
with integer coordinates in $\mathbb{R}^3 = \{ (x_1, x_2, x_3)|\ x_k
\in \mathbb{R},\  k =1, 2, 3 \}$ and fix initial data, for example,
on the coordinate axes of the lattice, $u_{i\, 0 0}$, $u_{ 0 j\,
0}$, and $u_{ 0 0 k}$, $i, j, k \in \mathbb{Z}$.


A two-dimensional discrete equation (\ref{1}) is said to be {\it
consistent on the cubic lattice} (see [2]--[6]) if for generic
initial data the discrete equation (\ref{1}) can be satisfied in a
consistent way simultaneously on all two-dimensional {\it
coordinate} sublattices of the cubic lattice $\mathbb{Z}^3$ that are
defined by fixing one of coordinates (any of the three coordinates)
of the cubic lattice. This condition is equivalent to the
consistency condition on each {\it elementary $2 \times 2 \times 2$
cube of the lattice $\mathbb{Z}^3$}, $\{(i + p, j + r, k + s), \ 0
\leq p, r, s \leq 1\}$, where $i, j,$ and $k$ are arbitrary fixed
integers, $i, j, k \in \mathbb{Z}$, i.e., relation (\ref{1}) must be
satisfied in a consistent way on all faces of any elementary $2
\times 2 \times 2$ cube of the lattice $\mathbb{Z}^3$ for generic
initial data. In the elementary cube $\{(i, j, k), \ 0 \leq i, j, k
\leq 1 \}$ the values $u_{1 0 1}$, $u_{1 1 0}$, and $u_{0 1 1}$ are
determined by relations (\ref{1}) on the corresponding faces of the
cube by the initial data $u_{0 0 0}$, $u_{1 0 0}$, $u_{0 1 0}$, and
$u_{0 0 1}$, and three relations on three faces of the cube must be
satisfied for the value $u_{1 1 1}$. One can consider the condition
of consistency of the overdetermined system of relations for the
value $u_{1 1 1}$ for generic initial data as the consistency
condition for the discrete equation (\ref{1}) on the cubic lattice
$\mathbb{Z}^3$.


Here, we do not discuss all various situations in which relations
(\ref{1}) correctly define a two-dimensional discrete equation for
the field $u$ on any two-dimensional coordinate sublattice of the
cubic lattice $\mathbb{Z}^3$; for example, one can assume for
simplicity that relations (\ref{1}) are invariant with respect to
the full symmetry group of the square. Classifications of discrete
equations of the form (\ref{1}) that are consistent on the cubic
lattice were studied in [5] and [9] under some additional conditions
on the functions $Q (x_1, x_2, x_3, x_4)$ (see also [10]). The
equation \be u_{i, j + 1} u_{i + 1, j} - u_{i + 1, j + 1} u_{i j} =
0, \ \ \ \ i, j \in \mathbb{Z}, \label{2}\ee defined by the
condition that the determinants of all $2\times2$ matrices of values
of the field $u$ at the vertices of elementary $2\times2$ squares of
the lattice $\mathbb Z^2$ vanish, is an example of such a
two-dimensional nonlinear discrete equation that is consistent on
the cubic lattice. Equation (\ref{2}) is linear with respect to each
variable and invariant with respect to the full symmetry group of
the square. Fixing arbitrary nonzero initial data $u_{i \, 0}$ and
$u_{ 0 j}$, $i, j \in \mathbb{Z}$, on the coordinate axes of the
lattice $\mathbb{Z}^2$ completely determines a field $u$ on the
lattice $\mathbb{Z}^2$ satisfying the discrete nonlinear equation
(\ref{2}), and fixing arbitrary nonzero initial data $u_{i \, 0 0}$,
$u_{ 0 j\, 0}$, and $u_{ 0 0 k}$, $i, j, k \in \mathbb{Z}$, on the
coordinate axes of the lattice $\mathbb{Z}^3$ completely determines
a field $u$ on the lattice $\mathbb{Z}^3$ satisfying the discrete
nonlinear equation (\ref{2}) on all two-dimensional coordinate
sublattices of the cubic lattice $\mathbb{Z}^3$; i.e., relations
(\ref{2}) are satisfied in a consistent way on all faces of each
elementary $2 \times 2 \times 2$ cube of the lattice $\mathbb{Z}^3$
for arbitrary nonzero initial data. The integrability (in the
broadest sense of the word) of the discrete nonlinear equation
(\ref{2}) is obvious, since it can be easily linearized: $ \ln u_{i,
j + 1} + \ln u_{i + 1, j} - \ln u_{i + 1, j + 1} - \ln u_{i j} = 0,
\ i, j \in \mathbb{Z}.$

In this article we consider the question of consistency on cubic
lattices for discrete nonlinear equations defined determinants of
arbitrary fixed orders $N > 2$. For $N > 2$ this question is
nontrivial, since the consistency condition on cubic lattices in the
form defined above is not satisfied in this case. We prove that
another consistency principle, {\it the modified consistency
principle on cubic lattices}, which was discovered in our paper [1],
holds for determinants of arbitrary orders $N > 2$.

\bigskip

\begin{center}
{\large \bf {Relations on elementary $3 \times 3$ squares\\ of the
lattice $\mathbb{Z}^2$ and consistency conditions}}
\end{center}

\medskip
We will use the following definition everywhere in this paper. An
{\it elementary $N \times N$ square of the square lattice
$\mathbb{Z}^2$} is a set of points of the lattice $\mathbb{Z}^2$
with coordinates $\{(i + s, j + r), \ 0 \leq s, r \leq N - 1\}$,
where $i, j$ is an arbitrary fixed pair of integers, $i, j \in
\mathbb{Z}$, $N \geq 2$.

Let us consider a discrete equation on $\mathbb{Z}^2$ defined by a
relation for the values of the field $u$ at the points of the
lattice $\mathbb{Z}^2$ that form {\it elementary $3 \times 3$
squares}: \be Q (u_{i j}, \ldots, u_{i + s, j + r}, \ldots, u_{i +
2, j + 2}) = 0, \ \  0 \leq s, r \leq 2, \ \ i, j \in \mathbb{Z},
\label{4}\ee so that in each elementary $3 \times 3$ square of the
lattice $\mathbb{Z}^2$, i.e., in each set of lattice points with
coordinates of the form $\{(i, j), (i + 1, j), (i + 2, j), (i, j +
1), (i + 1, j + 1), (i + 2, j + 1), (i, j + 2), (i + 1, j + 2), (i +
2, j + 2)\}$, $i, j \in \mathbb{Z}$, the value of the field $u$ at
one of the points of this elementary $3 \times 3$ square is
determined by the values of the field at the other eight points.

For definiteness, one can require, for example, that relations
(\ref{4}) are invariant with respect to the full symmetry group of
the configuration of points of the lattice $\mathbb{Z}^2$ that form
elementary $3 \times 3$ squares (obviously, this group of symmetries
coincides with the full symmetry group of the usual square). For any
discrete equation of the form (\ref{4}), fixing generic initial
data, for example, on two bands along the coordinate axes of the
lattice $\mathbb{Z}^2$, $\{(i, 0), (i, 1), i \in \mathbb{Z} \}$ and
$\{(0, j), (1, j), j \in \mathbb{Z} \}$, completely determines a
field $u$ on $\mathbb{Z}^2$ that satisfies this equation.

Quite similarly, discrete equations on $\mathbb{Z}^2$ given by
relations for the values of the field $u$ at the points of the
lattice $\mathbb{Z}^2$ that form elementary $N \times N$ squares can
be defined for an arbitrary $N \geq 2$. For definiteness, one can
again require, for example, that the relations are invariant with
respect to the full symmetry group of the configuration of points of
the lattice $\mathbb{Z}^2$ that form elementary $N \times N$ squares
(moreover, it is also obvious that for any $N \geq 2$ the full
symmetry group of the configuration of points of the lattice
$\mathbb{Z}^2$ that form elementary $N \times N$ squares coincides
with the full symmetry group of the usual square).

We consider the cubic lattice $\mathbb{Z}^3$ and {\it the
consistency condition for discrete equations of the form {\rm
(\ref{4})} on all two-dimensional coordinate sublattices of the
cubic lattice $\mathbb{Z}^3$}. Initial data can be specified, for
example, at the following lattice points that are situated on 12
straight lines going along the coordinate axes of the lattice
$\mathbb{Z}^3$: $(i, 0, 0)$, $(i, 1, 0)$, $(i, 0, 1)$, $(i, 1, 1)$,
$(0, j, 0)$, $(1, j, 0)$, $(0, j, 1)$, $(1, j, 1)$, $(0, 0, k)$,
$(1, 0, k)$, $(0, 1, k)$, and $(1, 1, k)$, $i, j, k \in \mathbb{Z}$.
The values of the field $u$ at all other points of the cubic lattice
$\mathbb{Z}^3$ must then be correctly determined in a consistent way
by relations (\ref{4}) on all elementary $3 \times 3$ squares of all
two-dimensional coordinate sublattices of the cubic lattice
$\mathbb{Z}^3$ for generic initial data. In the elementary cube
$\{(i, j, k), \ 0 \leq i, j, k \leq 2 \}$ the values $u_{2 0 2}$,
$u_{2 1 2}$, $u_{2 2 0}$, $u_{2 2 1}$, $u_{0 2 2}$, and $u_{1 2 2}$
are determined using relations (\ref{4}) on the corresponding
elementary $3 \times 3$ squares situated in the cube under
consideration (six faces and three middle normal sections of the
cube) by the initial data $u_{0 0 0}$, $u_{1 0 0}$, $u_{2 0 0}$,
$u_{0 1 0}$, $u_{1 1 0}$, $u_{2 1 0}$, $u_{0 0 1}$, $u_{1 0 1}$,
$u_{2 0 1}$, $u_{0 1 1}$, $u_{1 1 1}$, $u_{2 1 1}$, $u_{0 2 0}$,
$u_{1 2 0}$, $u_{0 2 1}$, $u_{1 2 1}$, $u_{0 0 2}$, $u_{1 0 2}$,
$u_{0 1 2}$, and $u_{1 1 2}$, and three relations must hold
simultaneously on three faces of the cube for the value $u_{2 2 2}$.
One can assume that the consistency condition on the cubic lattice
$\mathbb{Z}^3$ for any discrete equation of the form (\ref{4}) is
the consistency condition of the corresponding overdetermined system
of relations on the value $u_{2 2 2}$ for generic initial data.

Quite similarly, for an arbitrary  $N \geq 2$, one can define {\it
the consistency condition on the cubic lattice $\mathbb{Z}^3$ for
discrete equations on $\mathbb{Z}^2$ given by relations on the
values of the field $u$ at the points of the lattice $\mathbb{Z}^2$
that form elementary $N \times N$ squares}. The values of the field
$u$ at all points of the cubic lattice $\mathbb{Z}^3$ must be
correctly determined in a consistent way by relations on all
elementary $N \times N$ squares of all two-dimensional coordinate
sublattices of the cubic lattice $\mathbb{Z}^3$ for generic initial
data. The consistency of discrete equations on the lattice
$\mathbb{Z}^2$ that are given by relations on the values of the
field $u$ at the points of the lattice $\mathbb{Z}^2$ that form
elementary $N \times N$ squares can be considered on one elementary
$N \times N \times N$ cube (the consistency of the relations on all
faces and all normal sections of the cube that are parallel to the
coordinate planes for generic initial data specified in the cube).

Let us consider, in particular, the discrete nonlinear equation on
$\mathbb{Z}^2$ defined by the condition that the determinants of all
$3 \times 3$ matrices of values of the field $u$ at the points of
the lattice $\mathbb{Z}^2$ that form elementary $3 \times 3$ squares
vanish: \bea && u_{i, j + 2} u_{i + 1, j + 1} u_{i + 2, j} + u_{i, j
+ 1} u_{i + 1, j} u_{i + 2, j + 2} + u_{i, j} u_{i + 1, j +
2} u_{i + 2, j + 1} - \nn \\
&& - u_{i, j} u_{i + 1, j + 1} u_{i + 2, j + 2} - u_{i, j + 2} u_{i
+ 1, j} u_{i + 2, j + 1} - u_{i, j + 1} u_{i + 1, j + 2} u_{i + 2,
j} = 0, \ i, j \in \mathbb{Z}. \label{5}\eea Equation (\ref{5}) is
linear with respect to each variable and invariant with respect to
the full symmetry group of the configuration of points of the
lattice $\mathbb{Z}^2$ that form elementary $3 \times 3$ squares.

It is not difficult to check that for generic initial data the
above-considered consistency condition on the cubic lattice is not
satisfied for the discrete equation (\ref{5}), and, in this sense,
the discrete nonlinear equation (\ref{5}) is not consistent on
two-dimensional coordinate sublattices of the cubic lattice
$\mathbb{Z}^3$.

We note that for such setting of the consistency problem on the
cubic lattice for discrete equations of the form (\ref{4}) we have
the following: in the elementary $3 \times 3 \times 3$ cube $\{(i,
j, k), \ 0 \leq i, j, k \leq 2 \}$ of the lattice $\mathbb{Z}^3$,
given initial data of values of the field $u$ at 20 points of this
elementary $3 \times 3 \times 3$ cube, one can determine the values
of the field $u$ at the other seven points of this elementary $3
\times 3 \times 3$ cube by relations (\ref{4}) (nine relations on
six faces and on three middle normal sections of the cube), and only
for the value of the field at one of the points one obtains an
overdetermined system consisting of three relations on three
distinct elementary $3 \times 3$ squares.

In the paper [1] we proposed another consistency condition, {\it a
modified consistency condition on cubic lattices} for discrete
nonlinear equations of the form (\ref{4}) and their $N \times N$
generalizations, which works perfectly well at least for discrete
nonlinear equations given by determinants. We hope that this
approach will also be useful for other discrete equations, and we
plan to carry out further investigations in this direction,
including some classification problems and generalizations. In
particular, we hope very much that this approach and its natural
generalizations will be useful for the study of the consistency
principles for discrete nonlinear equations given by
hyperdeterminants, which requires serious programs of symbol
calculations (see the interesting paper by Tsarev and Wolf [10],
which was one of the most important stimuli to our investigations).

\bigskip

\begin{center}
{\large \bf {Bent elementary $3 \times 3$ squares\\
and modified consistency conditions}}
\end{center}

\medskip

We consider a discrete equation of the form (\ref{4}) and require
that the discrete equation is satisfied not only on all
two-dimensional coordinate sublattices of the cubic lattice
$\mathbb{Z}^3$, but also on all unions of two arbitrary intersecting
two-dimensional coordinate sublattices of the cubic lattice
$\mathbb{Z}^3$; i.e., the corresponding elementary $3 \times 3$
squares on which the discrete equation of the form (\ref{4}) is
considered can be {\it bent} at a right angle passing from one
two-dimensional coordinate sublattice to another, for example,
$\{(i, 0, 0), (i, 1, 0), (i, 0, 1), i = 0, 1, 2\}$, $\{(0, j, 0),
(1, j, 0), (0, j, 1), j = 0, 1, 2\}$, and $\{(0, 0, k), (1, 0, k),
(0, 1, k), k = 0, 1, 2\}$. In this case initial data can be
specified, for example, at the following points of the cubic lattice
$\mathbb{Z}^3$: $(i, 0, 0)$, $(i, 0, 1)$, $(2, j, 0)$, $(2, j, 1)$,
$(1, 0, k)$, $(2, 0, k)$, $(1, 1, 0)$, and $(1, 1, 1)$, $i, j, k \in
\mathbb{Z}$. The values of the field $u$ at all other points of the
cubic lattice $\mathbb{Z}^3$ must then be correctly determined in a
consistent way by relations (\ref{4}) on all elementary $3 \times 3$
squares (including all bent elementary $3 \times 3$ squares) of all
unions of two arbitrary intersecting two-dimensional coordinate
sublattices of the cubic lattice $\mathbb{Z}^3$ for generic initial
data. In the elementary cube $\{(i, j, k), \ 0 \leq i, j, k \leq 2
\}$ the values $u_{2 1 2}$, $u_{2 2 2}$, $u_{1 1 2}$, $u_{0 0 2}$,
$u_{1 2 k}$, $u_{0 1 k}$, and $u_{0 2 k}$, $0 \leq k \leq 2$, are
determined by the initial data $u_{0 0 0}$, $u_{1 0 0}$, $u_{2 0
0}$, $u_{0 0 1}$, $u_{1 1 0}$, $u_{2 1 0}$, $u_{1 0 1}$, $u_{2 0
1}$, $u_{1 1 1}$, $u_{2 1 1}$, $u_{2 0 2}$, $u_{2 2 0}$, $u_{2 2
1}$, and $u_{1 0 2}$ and by overdetermined systems generated by
relations (\ref{4}) on elementary $3 \times 3$ squares (including
all bent elementary $3 \times 3$ squares) that are situated in the
cube under consideration (six faces, three middle normal sections of
the cube, and 48 bent elementary $3 \times 3$ squares). There are 48
distinct bent elementary $3 \times 3$ squares in any elementary $3
\times 3 \times 3$ cube, which can be easily counted by the bending
edges of the bent elementary $3 \times 3$ squares: to each of the 12
edges of the cube, there corresponds one bent elementary $3 \times
3$ square in the cube; to each of the 12 middle lines of points on
the faces of the cube ($2 \times 6$), there correspond two distinct
bent elementary $3 \times 3$ squares in the cube; and to each of the
three interior lines of points in the cube that connect the centres
of opposite faces of the cube, there correspond four distinct bent
elementary $3 \times 3$ squares in the cube. One can assume that the
consistency condition on the cubic lattice $\mathbb{Z}^3$ for any
discrete equation of the form (\ref{4}) is the consistency condition
of the corresponding overdetermined system of relations on the
values of the field $u$ in the elementary cube $\{(i, j, k), \ 0
\leq i, j, k \leq 2 \}$ for generic initial data. The corresponding
discrete equations will also be called {\it consistent on the cubic
lattice}.

We note that for this new setting of the consistency problem on the
cubic lattice for discrete equations of the form (\ref{4}) we have
the following: in the elementary $3 \times 3 \times 3$ cube $\{(i,
j, k), \ 0 \leq i, j, k \leq 2 \}$ of the lattice $\mathbb{Z}^3$,
given initial data of values of the field $u$ at 14 points of this
elementary $3 \times 3 \times 3$ cube, one can determine the values
of the field $u$ at the other 13 points of this elementary $3 \times
3 \times 3$ cube by relations (\ref{4}) (57 relations on six faces,
on three middle normal sections of the cube, and on 48 bent
elementary $3 \times 3$ squares), which constitute in this case a
highly overdetermined system of relations.

Quite similarly, for an arbitrary $N > 2$, one can define the
corresponding {\it modified consistency condition on the cubic
lattice $\mathbb{Z}^3$ for discrete equations on the square lattice
$\mathbb{Z}^2$ that are given by relations on the values of the
field $u$ at the points of the lattice $\mathbb{Z}^2$ that form
elementary $N \times N$ squares {\rm (}including any bent elementary
$N \times N$ squares{\rm )}}. Moreover, for $N > 3$, one can,
generally speaking, allow a larger number (up to $N - 2$) of
bendings of elementary $N \times N$ squares in the cubic lattice
$\mathbb{Z}^3$ (in this case each elementary $N \times N$ square can
be bent in the cubic lattice simultaneously along up to $N - 2$
parallel lines of the same type, each bending being to one of the
two possible different sides).

The following basic theorem holds.

{\bf Theorem 1} [1]. {\it For arbitrary generic initial data, the
nonlinear discrete equation {\rm (\ref{5})} can be satisfied in a
consistent way on all unions of pairs of arbitrary intersecting
two-dimensional coordinate sublattices of the cubic lattice
$\mathbb{Z}^3$; i.e., the discrete nonlinear equation {\rm
(\ref{5})} is consistent on the cubic lattice $\mathbb{Z}^3$.}

{\it Proof.} Let us consider the elementary cube $\{(i, j, k), \ 0
\leq i, j, k \leq 2 \}$ of the cubic lattice $\mathbb{Z}^3$ and
specify generic initial data at the following points of this
elementary cube: $(i, 0, 0)$, $(i, 0, 1)$, $(2, j, 0)$, $(2, j, 1)$,
$(1, 0, k)$, $(2, 0, k)$, $(1, 1, 0)$, and $(1, 1, 1)$, $0 \leq i,
j, k \leq 2$. We will distinguish the following three different
types of lines of lattice points in the elementary cube under
consideration: the lines parallel to the $x$-axis, i.e., the sets of
points of the form $\{(i, r, s), 0 \leq i \leq 2\}$, where $(r, s)$
are fixed ordered pairs of integers, $0 \leq r, s \leq 2$, that
number the lines in this elementary cube that are parallel to the
$x$-axis ({\it the $x$-type lines}); the lines parallel to the
$y$-axis, i.e., the sets of points of the form $\{(r, j, s), 0 \leq
j \leq 2\}$, where $(r, s)$ are fixed ordered pairs of integers, $0
\leq r, s \leq 2$, that numb the lines in the elementary cube under
consideration that are parallel to the $y$-axis ({\it the $y$-type
lines}); and the lines parallel to the $z$-axis, i.e., the sets of
points of the form $\{(r, s, k), 0 \leq k \leq 2\}$, where $(r, s)$
are fixed ordered pairs of integers, $0 \leq r, s \leq 2$, that
number the lines in the elementary cube under consideration that are
parallel to the $z$-axis ({\it the $z$-type lines}). The specified
initial data fill a pair of lines of each of these three types (a
pair of lines parallel to the corresponding coordinate axis for each
of the coordinate axes). We will consider all these lines as {\it
basic} ones: $\{(i, 0, 0), 0 \leq i \leq 2\}$ and $\{(i, 0, 1), 0
\leq i \leq 2\}$ ({\it the basic $x$-type lines}); $\{(2, j, 0), 0
\leq j \leq 2\}$ and $\{(2, j, 1), 0 \leq j \leq 2\}$ ({\it the
basic $y$-type lines}); and $\{(1, 0, k), 0 \leq k \leq 2\}$ and
$\{(2, 0, k), 0 \leq k \leq 2\}$ ({\it the basic $z$-type lines}).
Given arbitrary generic initial data, we will determine the values
of the scalar field $u$ at the remaining points of the elementary
cube under consideration according to relations (\ref{5}) and mark
the points at which the values of the field have already been found.
We will also shade each line in this elementary cube if the vector
of values of the field $u$ at the points of this line is a linear
combination of the vectors of values of the field $u$ at the points
of the two basic lines of the same type (for the coordinates of
vectors of values of the field $u$ at the points of lines of the
same type, there is a natural ascending order of the respective
coordinate $x$, $y$ or $z$). First of all, in the elementary cube
under consideration we must mark all lattice points at which the
initial data are given and shade all basic lines of all three types
by the very definition of this procedure. It is obvious that if
carrying out such a procedure for generic initial data yields all
the lattice points marked and all the lines of all the types shaded
in the elementary cube under consideration, then the theorem will be
proved, because in this case, {\it for any three lines of the same
type in this elementary cube} (and, hence, for any elementary $3
\times 3$ square in this elementary cube, bent or unbent), the
determinant of the matrix of values of the scalar field $u$ at the
points of these lines will vanish, and this is even more than is
required for the consistency of the corresponding discrete equation.
Thus, in this case, as a matter of fact, we will prove even a {\it
considerably stronger principle of consistency on the cubic lattice
$\mathbb{Z}^3$ for determinants and for the nonlinear discrete
equation {\rm (\ref{5})}}. It remains to shade all the lines of all
the types in the elementary cube under consideration. For this
purpose, it is necessary to consider consecutively at least 13
elementary $3 \times 3$ squares (bent and unbent) of our elementary
cube.

Let us consider the elementary $3 \times 3$ square $\{(i, 0, 0), (i,
0, 1), (i, 0, 2), i = 0, 1, 2\}$ in our cube (a face of the cube).
In this elementary square the values of the field $u$ are given at
eight points and, consequently, the value of the field $u$ at the
remaining ninth point $(0, 0, 2)$ is determined by relation {\rm
(\ref{5})}, i.e., by the condition that the determinant of the
matrix of values of the field at the lattice points of this
elementary $3 \times 3$ square vanishes. Therefore, the vector of
values of the field $u$ at the points of the line $\{(i, 0, 2), 0
\leq i \leq 2\}$ is a linear combination of the vectors of values of
the field $u$ at the points of the two basic lines of the same type,
$\{(i, 0, 0), 0 \leq i \leq 2\}$ and $\{(i, 0, 1), 0 \leq i \leq
2\}$, situated in the given elementary $3 \times 3$ square; i.e., we
can mark the point $(0, 0, 2)$ and shade the line $\{(i, 0, 2), 0
\leq i \leq 2\}$.

Similarly, since the determinant of the matrix of values of the
field at the lattice points of this elementary $3 \times 3$ square
vanishes, it follows immediately that the vector of values of the
field $u$ at the points of the line $\{(0, 0, k), 0 \leq k \leq 2\}$
is a linear combination of the vectors of values of the field $u$ at
the points of the other two lines of this elementary $3 \times 3$
square, namely, the two basic lines of the same type, $\{(1, 0, k),
0 \leq k \leq 2\}$ and $\{(2, 0, k), 0 \leq k \leq 2\}$, situated in
the given elementary $3 \times 3$ square; i.e., we can shade the
line $\{(0, 0, k), 0 \leq k \leq 2\}$.

Let us consider the bent elementary $3 \times 3$ square $\{(1, 0,
k), (2, 0, k), (2, 1, k), k = 0, 1, 2\}$ in our cube. In this
elementary square the values of the field $u$ are given at eight
points and, consequently, the value of the field $u$ at the
remaining ninth point $(2, 1, 2)$ is determined by relation {\rm
(\ref{5})}, i.e., by the condition that the determinant of the
matrix of values of the field at the lattice points of this bent
elementary $3 \times 3$ square vanishes, which immediately implies
that the vector of values of the field $u$ at the points of the line
$\{(2, 1, k), 0 \leq k \leq 2\}$ is a linear combination of the
vectors of values of the field $u$ at the points of the two basic
lines of the same type, $\{(1, 0, k), 0 \leq k \leq 2\}$ and $\{(2,
0, k), 0 \leq k \leq 2\}$, situated in the given bent elementary $3
\times 3$ square; i.e., we can mark the point $(2, 1, 2)$ and shade
the line $\{(2, 1, k), 0 \leq k \leq 2\}$.

Now we consider another bent elementary $3 \times 3$ square $\{(1,
1, k), (1, 0, k), (2, 0, k), k = 0, 1, 2\}$ in our cube. In this
elementary square the values of the field $u$ are given at eight
points and, consequently, the value of the field $u$ at the
remaining ninth point $(1, 1, 2)$ is determined by relation {\rm
(\ref{5})}, i.e., by the condition that the determinant of the
matrix of values of the field at the lattice points of this bent
elementary $3 \times 3$ square vanishes, which immediately implies
that the vector of values of the field $u$ at the points of the line
$\{(1, 1, k), 0 \leq k \leq 2\}$ is a linear combination of the
vectors of values of the field $u$ at the points of the two basic
lines of the same type, $\{(1, 0, k), 0 \leq k \leq 2\}$ and $\{(2,
0, k), 0 \leq k \leq 2\}$, situated in the given bent elementary $3
\times 3$ square; i.e., we can mark the point $(1, 1, 2)$ and shade
the line $\{(1, 1, k), 0 \leq k \leq 2\}$.

Let us consider one more bent elementary $3 \times 3$ square $\{(i,
0, 0), (i, 0, 1), (i, 1, 1), i = 0, 1, 2\}$ in our cube. In this
elementary square the values of the field $u$ are given at eight
points and, consequently, the value of the field $u$ at the
remaining ninth point $(0, 1, 1)$ is determined by relation {\rm
(\ref{5})}, i.e., by the condition that the determinant of the
matrix of values of the field at the lattice points of this bent
elementary $3 \times 3$ square vanishes, which immediately implies
that the vector of values of the field $u$ at the points of the line
$\{(i, 1, 1), 0 \leq i \leq 2\}$ is a linear combination of the
vectors of values of the field $u$ at the points of the two basic
lines of the same type, $\{(i, 0, 0), 0 \leq i \leq 2\}$ and $\{(i,
0, 1), 0 \leq i \leq 2\}$, situated in the given bent elementary $3
\times 3$ square; i.e., we can mark the point $(0, 1, 1)$ and shade
the line $\{(i, 1, 1), 0 \leq i \leq 2\}$.

Let us consider the next bent elementary $3 \times 3$ square $\{(i,
0, 1), (i, 0, 0), (i, 1, 0),  i = 0, 1, 2\}$ in our cube. In this
elementary square the values of the field $u$ are given at eight
points and, consequently, the value of the field $u$ at the
remaining ninth point $(0, 1, 0)$ is determined by relation {\rm
(\ref{5})}, i.e., by the condition that the determinant of the
matrix of values of the field at the lattice points of this bent
elementary $3 \times 3$ square vanishes, which immediately implies
that the vector of values of the field $u$ at the points of the line
$\{(i, 1, 0), 0 \leq i \leq 2\}$ is a linear combination of the
vectors of values of the field $u$ at the points of the two basic
lines of the same type, $\{(i, 0, 0), 0 \leq i \leq 2\}$ and $\{(i,
0, 1), 0 \leq i \leq 2\}$, situated in the given bent elementary $3
\times 3$ square; i.e., we can mark the point $(0, 1, 0)$ and shade
the line $\{(i, 1, 0), 0 \leq i \leq 2\}$.

Let us consider one more bent elementary $3 \times 3$ square $\{(1,
j, 0), (2, j, 0), (2, j, 1),  j = 0, 1, 2\}$ in our cube. In this
elementary square the values of the field $u$ are given at eight
points and, consequently, the value of the field $u$ at the
remaining ninth point $(1, 2, 0)$ is determined by relation {\rm
(\ref{5})}, i.e., by the condition that the determinant of the
matrix of values of the field at the lattice points of this bent
elementary $3 \times 3$ square vanishes, which immediately implies
that the vector of values of the field $u$ at the points of the line
$\{(1, j, 0), 0 \leq j \leq 2\}$ is a linear combination of the
vectors of values of the field $u$ at the points of the two basic
lines of the same type, $\{(2, j, 0), 0 \leq j \leq 2\}$ and $\{(2,
j, 1), 0 \leq j \leq 2\}$, situated in the given bent elementary $3
\times 3$ square; i.e., we can mark the point $(1, 2, 0)$ and shade
the line $\{(1, j, 0), 0 \leq j \leq 2\}$.

Let us consider the next bent elementary $3 \times 3$ square $\{(2,
j, 0), (2, j, 1), (1, j, 1), j = 0, 1, 2\}$ in our cube. In this
elementary square the values of the field $u$ are given at eight
points and, consequently, the value of the field $u$ at the
remaining ninth point $(1, 2, 1)$ is determined by relation {\rm
(\ref{5})}, i.e., by the condition that the determinant of the
matrix of values of the field at the lattice points of this bent
elementary $3 \times 3$ square vanishes, which immediately implies
that the vector of values of the field $u$ at the points of the line
$\{(1, j, 1), 0 \leq j \leq 2\}$ is a linear combination of the
vectors of values of the field $u$ at the points of the two basic
lines of the same type, $\{(2, j, 0), 0 \leq j \leq 2\}$ and $\{(2,
j, 1), 0 \leq j \leq 2\}$, situated in the given bent elementary $3
\times 3$ square; i.e., we can mark the point $(1, 2, 1)$ and shade
the line $\{(1, j, 1), 0 \leq j \leq 2\}$.

Let us consider one more elementary $3 \times 3$ square $\{(2, j,
0), (2, j, 1), (2, j, 2), j = 0, 1, 2\}$ in our cube (a face of the
cube). In this elementary square at the present moment the values of
the field $u$ are already determined at eight points and,
consequently, the value of the field $u$ at the remaining ninth
point $(2, 2, 2)$ is determined by relation {\rm (\ref{5})}, i.e.,
by the condition that the determinant of the matrix of values of the
field at the lattice points of this elementary $3 \times 3$ square
vanishes, which immediately implies that the vector of values of the
field $u$ at the points of the line $\{(2, j, 2), 0 \leq j \leq 2\}$
is a linear combination of the vectors of values of the field $u$ at
the points of the two basic lines of the same type, $\{(2, j, 0), 0
\leq j \leq 2\}$ and $\{(2, j, 1), 0 \leq j \leq 2\}$, situated in
the given elementary $3 \times 3$ square; i.e., we can mark the
point $(2, 2, 2)$ and shade the line $\{(2, j, 2), 0 \leq j \leq
2\}$.

We note that if the vector of values of the field $u$ at the lattice
points of an arbitrary line is a linear combination of the vectors
of values of the field $u$ at the lattice points of two shaded lines
of the same type, then this vector is a linear combination of the
vectors of values of the field $u$ at the points of the two basic
lines of the same type. This follows immediately from the fact that
each vector of values of the field $u$ at the points of an arbitrary
shaded line is a linear combination of the vectors of values of the
field $u$ at the points of the two basic lines of the same type.

Since the determinant of the matrix of values of the field at the
points of the elementary $3 \times 3$ square $\{(2, j, 0), (2, j,
1), (2, j, 2), j = 0, 1, 2\}$ (on a face of our cube) vanishes, it
follows immediately that the vector of values of the field $u$ at
the points of the line $\{(2, 2, k), 0 \leq k \leq 2\}$ is a linear
combination of the vectors of values of the field $u$ at the points
of two other lines of this elementary $3 \times 3$ square, namely,
two shaded lines of the same type, $\{(2, 0, k), 0 \leq k \leq 2\}$
and $\{(2, 1, k), 0 \leq k \leq 2\}$, situated in the given
elementary $3 \times 3$ square; i.e., we can shade the line $\{(2,
2, k), 0 \leq k \leq 2\}$.

In the elementary $3 \times 3$ square $\{(i, 0, 0), (i, 1, 0), (i,
2, 0), i = 0, 1, 2\}$ of our cube (on a face of the cube) at the
present moment the values of the field $u$ are already determined at
eight points and, consequently, the value of the field $u$ at the
remaining ninth point $(0, 2, 0)$ is determined by relation {\rm
(\ref{5})}, i.e., by the condition that the determinant of the
matrix of values of the field at the points of this elementary $3
\times 3$ square vanishes, which immediately implies that the vector
of values of the field $u$ at the points of the line $\{(i, 2, 0), 0
\leq i \leq 2\}$ is a linear combination of the vectors of values of
the field $u$ at the points of two shaded lines of the same type,
$\{(i, 0, 0), 0 \leq i \leq 2\}$ and $\{(i, 1, 0), 0 \leq i \leq
2\}$, situated in the given elementary $3 \times 3$ square; i.e., we
can mark the point $(0, 2, 0)$ and shade the line $\{(i, 2, 0), 0
\leq i \leq 2\}$.

Since the determinant of the matrix of values of the field at the
points of the elementary $3 \times 3$ square $\{(i, 0, 0), (i, 1,
0), (i, 2, 0), i = 0, 1, 2\}$ (on a face of our cube) vanishes, it
follows immediately that the vector of values of the field $u$ at
the points of the line $\{(0, j, 0), 0 \leq j \leq 2\}$ is a linear
combination of the vectors of values of the field $u$ at the points
of two other lines of this elementary $3 \times 3$ square, namely,
two shaded lines of the same type, $\{(1, j, 0), 0 \leq j \leq 2\}$
and $\{(2, j, 0), 0 \leq j \leq 2\}$, situated in the given
elementary $3 \times 3$ square; i.e., we can shade the line $\{(0,
j, 0), 0 \leq j \leq 2\}$.

In the elementary $3 \times 3$ square $\{(i, 0, 1), (i, 1, 1), (i,
2, 1), i = 0, 1, 2\}$ of our cube (on a middle normal section of the
cube) at the present moment the values of the field $u$ are already
determined at eight points and, consequently, the value of the field
$u$ at the remaining ninth point $(0, 2, 1)$ is determined by
relation {\rm (\ref{5})}, i.e., by the condition that the
determinant of the matrix of values of the field at the points of
this elementary $3 \times 3$ square vanishes, which immediately
implies that the vector of values of the field $u$ at the points of
the line $\{(i, 2, 1), 0 \leq i \leq 2\}$ is a linear combination of
the vectors of values of the field $u$ at the points of two shaded
lines of the same type, $\{(i, 0, 1), 0 \leq i \leq 2\}$ and $\{(i,
1, 1), 0 \leq i \leq 2\}$, situated in the given elementary $3
\times 3$ square; i.e., we can mark the point $(0, 2, 1)$ and shade
the line $\{(i, 2, 1), 0 \leq i \leq 2\}$.

Since the determinant of the matrix of values of the field at the
points of the elementary $3 \times 3$ square $\{(i, 0, 1), (i, 1,
1), (i, 2, 1), i = 0, 1, 2\}$ (on a middle normal section of our
cube) vanishes, it follows immediately that the vector of values of
the field $u$ at the points of the line $\{(0, j, 1), 0 \leq j \leq
2\}$ is a linear combination of the vectors of values of the field
$u$ at the points of two other lines of this elementary $3 \times 3$
square, namely, two shaded lines of the same type, $\{(1, j, 1), 0
\leq j \leq 2\}$ and $\{(2, j, 1), 0 \leq j \leq 2\}$, situated in
the given elementary $3 \times 3$ square; i.e., we can shade the
line $\{(0, j, 1), 0 \leq j \leq 2\}$.

Let us consider one more elementary $3 \times 3$ square $\{(1, j,
0), (1, j, 1), (1, j, 2), j = 0, 1, 2\}$ in our cube (a middle
normal section of the cube). In this elementary square at the
present moment the values of the field $u$ are already determined at
eight points and, consequently, the value of the field $u$ at the
remaining ninth point $(1, 2, 2)$ is determined by relation {\rm
(\ref{5})}, i.e., by the condition that the determinant of the
matrix of values of the field at the points of this elementary $3
\times 3$ square vanishes, which immediately implies that the vector
of values of the field $u$ at the points of the line $\{(1, j, 2), 0
\leq j \leq 2\}$ is a linear combination of the vectors of values of
the field $u$ at the points of two shaded lines of the same type,
$\{(1, j, 0), 0 \leq j \leq 2\}$ and $\{(1, j, 1), 0 \leq j \leq
2\}$, situated in the given elementary $3 \times 3$ square; i.e., we
can mark the point $(1, 2, 2)$ and shade the line $\{(1, j, 2), 0
\leq j \leq 2\}$.

Since the determinant of the matrix of values of the field at the
points of the elementary $3 \times 3$ square $\{(1, j, 0), (1, j,
1), (1, j, 2), j = 0, 1, 2\}$ (on a middle normal section of our
cube) vanishes, it follows immediately that the vector of values of
the field $u$ at the points of the line $\{(1, 2, k), 0 \leq k \leq
2\}$ is a linear combination of the vectors of values of the field
$u$ at the points of two other lines of this elementary $3 \times 3$
square, namely, two shaded lines of the same type, $\{(1, 0, k), 0
\leq k \leq 2\}$ and $\{(1, 1, k), 0 \leq k \leq 2\}$, situated in
the given elementary $3 \times 3$ square; i.e., we can shade the
line $\{(1, 2, k), 0 \leq k \leq 2\}$.

Let us consider one more elementary $3 \times 3$ square $\{(i, 1,
0), (i, 1, 1), (i, 1, 2), i = 0, 1, 2\}$ in our cube (a middle
normal section of the cube). In this elementary square at the
present moment the values of the field $u$ are already determined at
eight points and, consequently, the value of the field $u$ at the
remaining ninth point $(0, 1, 2)$ is determined by relation {\rm
(\ref{5})}, i.e., by the condition that the determinant of the
matrix of values of the field at the points of this elementary $3
\times 3$ square vanishes. Hence the vector of values of the field
$u$ at the points of the line $\{(i, 1, 2), 0 \leq i \leq 2\}$ is a
linear combination of the vectors of values of the field $u$ at the
points of two shaded lines of the same type, $\{(i, 1, 0), 0 \leq i
\leq 2\}$ and $\{(i, 1, 1), 0 \leq i \leq 2\}$, situated in the
given elementary $3 \times 3$ square; i.e., we can mark the point
$(0, 1, 2)$ and shade the line $\{(i, 1, 2), 0 \leq i \leq 2\}$.

Since the determinant of the matrix of values of the field at the
points of the elementary $3 \times 3$ square $\{(i, 1, 0), (i, 1,
1), (i, 1, 2), i = 0, 1, 2\}$ (on a middle normal section of our
cube) vanishes, it follows immediately that the vector of values of
the field $u$ at the points of the line $\{(0, 1, k), 0 \leq k \leq
2\}$ is a linear combination of the vectors of values of the field
$u$ at the points of two other lines of this elementary $3 \times 3$
square, namely, two shaded lines of the same type, $\{(1, 1, k), 0
\leq k \leq 2\}$ and $\{(2, 1, k), 0 \leq k \leq 2\}$, situated in
the given elementary $3 \times 3$ square; i.e., we can shade the
line $\{(0, 1, k), 0 \leq k \leq 2\}$.

It remains to determine the value of the field $u$ only at one point
of our cube, and three edges of the cube are still unshaded for the
present.

Let us consider the elementary $3 \times 3$ square $\{(i, 2, 0), (i,
2, 1), (i, 2, 2), i = 0, 1, 2\}$ in our cube (a face of the cube).
In this elementary square at the present moment the values of the
field $u$ are already determined at eight points and, consequently,
the value of the field $u$ at the remaining ninth point $(0, 2, 2)$
is determined by relation {\rm (\ref{5})}, i.e., by the condition
that the determinant of the matrix of values of the field at the
points of this elementary $3 \times 3$ square vanishes. Therefore,
the vector of values of the field $u$ at the points of the line
$\{(i, 2, 2), 0 \leq i \leq 2\}$ is a linear combination of the
vectors of values of the field $u$ at the points of two shaded lines
of the same type, $\{(i, 2, 0), 0 \leq i \leq 2\}$ and $\{(i, 2, 1),
0 \leq i \leq 2\}$, situated in the given elementary $3 \times 3$
square; i.e., we can mark the point $(0, 2, 2)$ and shade the line
$\{(i, 2, 2), 0 \leq i \leq 2\}$.

Now the values of the field $u$ are determined already at all points
of our cube, and it remains to shade two edges of the cube.

Since the determinant of the matrix of values of the field at the
points of the elementary $3 \times 3$ square $\{(i, 2, 0), (i, 2,
1), (i, 2, 2), i = 0, 1, 2\}$ (on a face of our cube) vanishes, it
follows immediately that the vector of values of the field $u$ at
the points of the line $\{(0, 2, k), 0 \leq k \leq 2\}$ is a linear
combination of the vectors of values of the field $u$ at the points
of two other lines of this elementary $3 \times 3$ square, namely,
two shaded lines of the same type, $\{(1, 2, k), 0 \leq k \leq 2\}$
and $\{(2, 2, k), 0 \leq k \leq 2\}$, situated in the given
elementary $3 \times 3$ square; i.e., we can shade the line $\{(0,
2, k), 0 \leq k \leq 2\}$.

Let us consider one more elementary $3 \times 3$ square $\{(0, j,
0), (0, j, 1), (0, j, 2), j = 0, 1, 2\}$ in our cube (a face of the
cube). In this elementary square at the present moment the values of
the field $u$ are already determined at all points, and the three
lines $\{(0, 0, k), 0 \leq k \leq 2\}$, $\{(0, 1, k), 0 \leq k \leq
2\}$, and $\{(0, 2, k), 0 \leq k \leq 2\}$ are shaded; i.e., the
vector of values of the field $u$ at the points of each of these
lines is a linear combination of the vectors of values of the field
$u$ at the points of the two basic lines of the same type. Thus, the
determinant of the matrix of values of the scalar field $u$ at the
points of the three lines $\{(0, 0, k), 0 \leq k \leq 2\}$, $\{(0,
1, k), 0 \leq k \leq 2\}$, and $\{(0, 2, k), 0 \leq k \leq 2\}$
vanishes. Hence, relation {\rm (\ref{5})} holds on the face $\{(0,
j, 0), (0, j, 1), (0, j, 2), j = 0, 1, 2\}$, and since the
determinant of the matrix of values of the field $u$ at the lattice
points of this face vanishes, it follows immediately that the vector
of values of the field $u$ at the points of the line $\{(0, j, 2), 0
\leq j \leq 2\}$ is a linear combination of the vectors of values of
the field $u$ at the points of two shaded lines of the same type,
$\{(0, j, 0), 0 \leq j \leq 2\}$ and $\{(0, j, 1), 0 \leq j \leq
2\}$, situated in the given elementary $3 \times 3$ square; i.e., we
can shade the line $\{(0, j, 2), 0 \leq j \leq 2\}$.

Thus, the values of the field $u$ are determined at all points of
our cube, and all lines of the cube are shaded now. The theorem is
proved.

Moreover, we have proved a {\it considerably stronger principle of
consistency on the cubic lattice for determinants}.

{\bf Theorem 2}. {\it For arbitrary generic initial data, the
nonlinear discrete equation {\rm (\ref{5})} can be satisfied in a
consistent way and simultaneously on each set of points of three
lines $P_l$, $1 \leq l \leq 3$, of the cubic lattice $\mathbb{Z}^3$
of the form $P_l = \{(i, r_l, s_l), a \leq i \leq a + 2\}$, $1 \leq
l \leq 3$, where $a$, $r_l$, and $s_l$, $1 \leq l \leq 3$, are
arbitrary fixed integers {\rm (}$x$-type lines{\rm )}, as well as on
each set of points of three lines $Q_l$, $1 \leq l \leq 3$, of the
cubic lattice $\mathbb{Z}^3$ of the form $Q_l = \{(r_l, j, s_l), a
\leq j \leq a + 2\}$, $1 \leq l \leq 3$, where $a$, $r_l$, and
$s_l$, $1 \leq l \leq 3$, are arbitrary fixed integers {\rm
(}$y$-type lines{\rm )}, and on each set of points of three lines
$R_l$, $1 \leq l \leq 3$, of the cubic lattice $\mathbb{Z}^3$ of the
form $R_l = \{(r_l, s_l, k), a \leq k \leq a + 2\}$, $1 \leq l \leq
3$, where $a$, $r_l$, and $s_l$, $1 \leq l \leq 3$, are arbitrary
fixed integers {\rm (}$z$-type lines{\rm )}. Moreover, in this case
the discrete equation {\rm (\ref{5})} will be satisfied in a
consistent way and simultaneously on each set of points of special
form lying on three bent lines $S_l$, $1 \leq l \leq 3$, of the same
type in the cubic lattice $\mathbb{Z}^3$, for example, of the form
$S_l = \{(a, r_l, s), (a + 1, r_l, s), (a + 1, r_l, s + 1)\}$, $1
\leq l \leq 3$, where $a$, $r_l$, and $s$, $1 \leq l \leq 3$, are
arbitrary fixed integers, of the form $S_l = \{(a, s, r_l), (a + 1,
s, r_l), (a + 1, s + 1, r_l)\}$, $1 \leq l \leq 3$, where $a$,
$r_l$, and $s$, $1 \leq l \leq 3$, are arbitrary fixed integers, or
of the form $S_l = \{(r_l, s, a + 1), (r_l, s, a), (r_l, s + 1,
a)\}$, $1 \leq l \leq 3$, where $a$, $r_l$, and $s$, $1 \leq l \leq
3$, are arbitrary fixed integers.}

The following {\it principle of consistency on the cubic lattice for
determinants} also holds.

Let us consider an arbitrary line $P$ (bent or unbent) given by
three arbitrary neighboring points in the cubic lattice
$\mathbb{Z}^3$. We consider an arbitrary set of three lines of the
cubic lattice $\mathbb{Z}^3$ that are obtained from the line $P$ by
translations in the lattice by vectors parallel to the
(one-dimensional or two-dimensional) space orthogonal to the line
$P$ (i.e., orthogonal to the plane or to the straight line of $P$
depending on whether the line $P$ is bent or unbent). Then, for
arbitrary generic initial data, the nonlinear discrete equation {\rm
(\ref{5})} can be satisfied in a consistent way and simultaneously
on each such set of three lines of the cubic lattice $\mathbb{Z}^3$.

Similar properties of consistency on cubic lattices hold for
determinants of arbitrary order $N \geq 2$.

{\bf Theorem 3} [1]. {\it For an arbitrary given positive integer $N
> 1$, the discrete nonlinear equation defined on the square lattice
$\mathbb{Z}^2$ by the condition that the determinants of all
matrices of values of the field $u$ at the points of the lattice
$\mathbb{Z}^2$ that form elementary $N \times N$ squares vanish is
consistent on the cubic lattice $\mathbb{Z}^3$.}

The proof for the general case of an arbitrary $N \geq 3$ is
completely similar to the case $N = 3$. It is important to note that
the analytical proof proposed in the present paper remains valid for
any positive integer $N > 1$ (for the discrete nonlinear equations
defined by the condition that the determinants of all matrices of
values of the field $u$ at the points of the lattice $\mathbb{Z}^2$
that form elementary $N \times N$ squares vanish). It is obvious
that after some consistency principle on the cubic lattice is
formulated, it can, of course, be easily checked by direct
calculations, for example, by an appropriate program of symbol
calculations (in this case the corresponding volume of calculations
and restricted possibilities of programs of symbol calculations may
be a serious problem), for any fixed $N$ and for any specific
discrete equation defined on elementary $N \times N$ squares;
however, it is clear that a consistency principle cannot be checked
for all $N$ by any calculations. For this purpose, we need an
analytical proof.

In particular, such an analytical proof can also be easily carried
out in the case $N = 2$ for the discrete nonlinear equation
(\ref{2}) without resorting to any calculations at all.

Indeed, let us consider the elementary cube $\{(i, j, k), \ 0 \leq
i, j, k \leq 1 \}$ of the cubic lattice $\mathbb{Z}^3$ and specify
generic initial data at the following points of this elementary
cube: $(i, 0, 0)$, $(0, j, 0)$, and $(0, 0, k)$, $0 \leq i, j, k
\leq 1$. As before, we will distinguish the following three
different types of lines of points in the elementary cube under
consideration: the lines parallel to the $x$-axis, i.e., the sets of
points of the form $\{(i, r, s), 0 \leq i \leq 1\}$, where $(r, s)$
are fixed ordered pairs of integers, $0 \leq r, s \leq 1$, that
number the lines in the elementary cube under consideration that are
parallel to the $x$-axis ({\it $x$-type lines}); the lines parallel
to the $y$-axis, i.e., the sets of points of the form $\{(r, j, s),
0 \leq j \leq 1\}$, where $(r, s)$ are fixed ordered pairs of
integers, $0 \leq r, s \leq 1$, that number the lines in the
elementary cube under consideration that are parallel to the
$y$-axis ({\it $y$-type lines}); and the lines parallel to the
$z$-axis, i.e., the sets of points of the form $\{(r, s, k), 0 \leq
k \leq 1\}$, where $(r, s)$ are fixed ordered pairs of integers, $0
\leq r, s \leq 1$, that number the lines in the elementary cube
under consideration that are parallel to the $z$-axis ({\it $z$-type
lines}). The specified initial data fill one line of each of these
three types (one of the lines that are parallel to the corresponding
coordinate axis for each of the coordinate axes). We will consider
all these lines as {\it basic} ones: $\{(i, 0, 0), 0 \leq i \leq
1\}$ ({\it the basic $x$-type lines}), $\{(0, j, 0), 0 \leq j \leq
1\}$ ({\it the basic $y$-type lines}), and $\{(0, 0, k), 0 \leq k
\leq 1\}$ ({\it the basic $z$-type lines}). Given arbitrary generic
initial data, we will determine the values of the scalar field $u$
at the remaining points of the elementary cube under consideration
according to relations (\ref{2}) and mark the points at which the
values of the field have already been found. We will also shade each
line in this elementary cube if the vector of values of the field
$u$ at the lattice points of this line is collinear with the vector
of values of the field $u$ at the lattice points of the basic line
of the same type (for the coordinates of vectors of values of the
field $u$ at the points of lines of the same type, there is a
natural ascending order of the respective coordinate $x$, $y$, or
$z$). First of all, in the elementary cube under consideration we
must mark all points at which the initial data are given and shade
all basic lines of all three types by the very definition itself of
this procedure. It is obvious that if carrying out such a procedure
for generic initial data yields all the lattice points marked and
all the lines of all the types shaded in the elementary cube under
consideration, then the consistency will be proved, because in this
case, {\it for any two lines of the same type in the elementary cube
under consideration} (and hence for any face of the elementary cube
under consideration), the determinant of the matrix of values of the
scalar field $u$ at the points of these lines will vanish, and this
is even more than is required for the consistency of the
corresponding discrete equation. Thus, in this case, as a matter of
fact, we will prove even a {\it stronger principle of consistency on
the cubic lattice $\mathbb{Z}^3$ for determinants and for the
nonlinear discrete equation {\rm (\ref{2})}}. It remains to shade
all the lines of all the types in the elementary cube under
consideration. To this end, it is necessary to consider
consecutively all faces  of our elementary cube.

Let us consider the face $\{(i, 0, 0), (i, 0, 1),  i = 0, 1\}$ in
our cube. On this face the values of the field $u$ are given at
three points and, consequently, the value of the field $u$ at the
remaining fourth point $(1, 0, 1)$ is determined by relation
(\ref{2}), i.e., by the condition that the determinant of the matrix
of values of the field at the lattice points of this face of the
cube vanishes. Therefore, the vector of values of the field $u$ at
the points of the line $\{(i, 0, 1), 0 \leq i \leq 1\}$ is collinear
with the vector of values of the field $u$ at the points of the
basic line of the same type $\{(i, 0, 0), 0 \leq i \leq 1\}$
situated on the given face; i.e., we can mark the point $(1, 0, 1)$
and shade the line $\{(i, 0, 1), 0 \leq i \leq 1\}$.

Similarly, since the determinant of the matrix of values of the
field $u$ at the lattice points on this face vanishes, it follows
immediately that the vector of values of the field $u$ at the points
of the line $\{(1, 0, k), 0 \leq k \leq 1\}$ is collinear with the
vector of values of the field $u$ at the points of another line of
this face, namely, the basic line of the same type $\{(0, 0, k), 0
\leq k \leq 1\}$ situated on the given face; i.e., we can shade the
line $\{(1, 0, k), 0 \leq k \leq 1\}$.

Let us consider the face $\{(0, j, 0), (0, j, 1), j = 0, 1\}$ in our
cube. On this face the values of the field $u$ are given at three
points and, consequently, the value of the field $u$ at the
remaining fourth point $(0, 1, 1)$ is determined by relation
(\ref{2}), i.e., by the condition that the determinant of the matrix
of values of the field at the points of this face of the cube
vanishes. Therefore, the vector of values of the field $u$ at the
points of the line $\{(0, j, 1), 0 \leq j \leq 1\}$ is collinear
with the vector of values of the field $u$ at the points of the
basic line of the same type $\{(0, j, 0), 0 \leq j \leq 1\}$
situated on the given face; i.e., we can mark the point $(0, 1, 1)$
and shade the line $\{(0, j, 1), 0 \leq j \leq 1\}$.

Since the determinant of the matrix of values of the field $u$ at
the points on the face $\{(0, j, 0), (0, j, 1),  j = 0, 1\}$ in our
cube vanishes, it follows immediately that the vector of values of
the field $u$ at the points of the line $\{(0, 1, k), 0 \leq k \leq
1\}$ is collinear with the vector of values of the field $u$ at the
points of another line of this face, namely, the basic line of the
same type $\{(0, 0, k), 0 \leq k \leq 1\}$ situated on the given
face; i.e., we can shade the line $\{(0, 1, k), 0 \leq k \leq 1\}$.

Let us consider the face $\{(0, j, 0), (1, j, 0), j = 0, 1\}$ in our
cube. On this face the values of the field $u$ are given at three
points and, consequently, the value of the field $u$ at the
remaining fourth point $(1, 1, 0)$ is determined by relation
(\ref{2}), i.e., by the condition that the determinant of the matrix
of values of the field at the points of this face of the cube
vanishes. Hence the vector of values of the field $u$ at the points
of the line $\{(1, j, 0), 0 \leq j \leq 1\}$ is collinear with the
vector of values of the field $u$ at the points of the basic line of
the same type $\{(0, j, 0), 0 \leq j \leq 1\}$ situated on the given
face; i.e., we can mark the point $(1, 1, 0)$ and shade the line
$\{(1, j, 0), 0 \leq j \leq 1\}$.

Since the determinant of the matrix of values of the field $u$ at
the points on the face $\{(0, j, 0), (1, j, 0),  j = 0, 1\}$ in our
cube vanishes, it follows immediately that the vector of values of
the field $u$ at the points of the line $\{(i, 1, 0), 0 \leq i \leq
1\}$ is collinear with the vector of values of the field $u$ at the
points of another line of this face, namely, the basic line of the
same type $\{(i, 0, 0), 0 \leq i \leq 1\}$ situated on the given
face; i.e., we can shade the line $\{(i, 1, 0), 0 \leq i \leq 1\}$.

It remains to determine the value of the field $u$ only at one point
of our cube, and three edges of the cube are still unshaded for the
present.

Let us consider the face $\{(1, j, 0), (1, j, 1), j = 0, 1\}$ in our
cube. On this face at the present moment the values of the field $u$
are already determined at three points and, consequently, the value
of the field $u$ at the remaining fourth point $(1, 1, 1)$ is
determined by relation (\ref{2}), i.e., by the condition that the
determinant of the matrix of values of the field at the points of
this face of the cube vanishes. Therefore, the vector of values of
the field $u$ at the points of the line $\{(1, j, 1), 0 \leq j \leq
1\}$ is collinear with the vector of values of the field $u$ at the
points of the shaded line of the same type $\{(1, j, 0), 0 \leq j
\leq 1\}$ situated on the given face, and then it is collinear with
the vector of values of the field $u$ at the points of the basic
line of the same type $\{(0, j, 0), 0 \leq j \leq 1\}$; i.e., we can
mark the point $(1, 1, 1)$ and shade the line $\{(1, j, 1), 0 \leq j
\leq 1\}$.

Now the values of the field $u$ are determined already at all
lattice points of our cube, and it remains to shade two edges of the
cube.

Since the determinant of the matrix of values of the field $u$ at
the points on the face $\{(1, j, 0), (1, j, 1),  j = 0, 1\}$ in our
cube vanishes, it follows immediately that the vector of values of
the field $u$ at the points of the line $\{(1, 1, k), 0 \leq k \leq
1\}$ is collinear with the vector of values of the field $u$ at the
points of another line of this face, namely, the shaded line of the
same type $\{(1, 0, k), 0 \leq k \leq 1\}$ situated on the given
face, and then it is collinear with the vector of values of the
field $u$ at the points of the basic line of the same type $\{(0, 0,
k), 0 \leq k \leq 1\}$; i.e., we can shade the line $\{(1, 1, k), 0
\leq k \leq 1\}$.

Let us consider the face $\{(i, 1, 0), (i, 1, 1),  i = 0, 1\}$ in
our cube. On this face at the present moment the values of the field
$u$ are already determined at all points, and two lines $\{(0, 1,
k), 0 \leq k \leq 1\}$ and $\{(1, 1, k), 0 \leq k \leq 1\}$ are
shaded; i.e., the vector of values of the field $u$ at the points of
each of these lines is collinear with the vector of values of the
field $u$ at the points of the basic line of the same type. Thus,
the determinant of the matrix of values of the scalar field $u$ at
the points of these two lines $\{(0, 1, k), 0 \leq k \leq 1\}$ and
$\{(1, 1, k), 0 \leq k \leq 1\}$ vanishes. Therefore, relation
(\ref{2}) holds on the face $\{(i, 1, 0), (i, 1, 1), i = 0, 1\}$,
and since the determinant of the matrix of values of the field at
the points on this face vanishes, it follows immediately that the
vector of values of the field $u$ at the points of the line $\{(i,
1, 1), 0 \leq i \leq 1\}$ is collinear with the vector of values of
the field $u$ at the points of the shaded line of the same type
$\{(i, 1, 0), 0 \leq i \leq 1\}$ situated on the given face, and
then it is collinear with the vector of values of the field $u$ at
the points of the basic line of the same type $\{(i, 0, 0), 0 \leq i
\leq 1\}$; i.e., we can shade the line $\{(i, 1, 1), 0 \leq i \leq
1\}$.

Thus, the values of the field $u$ are determined at all lattice
points of our cube, and all lines of the cube are shaded now. The
consistency is proved.

The proposed approach can also be applied to discrete equations
given on $N$-dimensi\-onal lattices and consistent on $(N +
1)$-dimensional lattices for arbitrary $N$. In particular, we hope
that it will be efficient for hyperdeterminants [14] and their
generalizations, as well as for other discrete equations on
$N$-dimensional lattices and other consistency principles.

It is also interesting to analyze the possibility of introducing
spectral parameters in the nonlinear discrete equations under
consideration that are consistent on cubic lattices, and to study
Lax pairs for them and other important properties of integrability.

In addition, it is very interesting to study various cases of {\it
partial consistency on $(N + 1)$-dimensional lattices for discrete
equations given on $N$-dimensional lattices}.

\medskip

{\bf {Acknowledgements.}} This paper was completed during the
author's visit to Taiwan, and the author is glad to express his
gratitude for the invitation and hospitality to Prof. Jen-Hsu Chang
and the National Defense University of Taiwan, to Prof. Jyh-Hao Lee,
to the Institute of Mathematics, Academia Sinica (Taipei, Taiwan),
and to the National Taiwan University (Taipei, Taiwan).

The work was carried out under partial financial support from the
Russian Foundation for Basic Research (project no.~09-01-00762),
from Siberian Federal University (grant $N^o$~26) and from the
programme ``Leading Scientific Schools'' (project no.
NSh-1824.2008.1).

\bigskip

\begin{center}
\bf {References}
\end{center}

\medskip

[1] O.I. Mokhov. On consistency of determinants on cubic lattices.
{\it Uspekhi Mat. Nauk}. 2008. V. {\bf 63}, No. 6. P. 169--170;
English translation in {\it Russian Math. Surveys}. 2008. V. {\bf
63}, No. 6. P. 1146--1148; http://arxiv.org/abs/0809/2032.

[2] F.W. Nijhoff, A.J. Walker. The discrete and continuous
Painlev\'e VI hierarchy and the Garnier systems. Integrable systems:
linear and nonlinear dynamics (Islay, 1999). {\it Glasgow Math. J.},
{\bf 43A} (2001), 109--123; arXiv:nlin/0001054.

[3] F.W. Nijhoff. Lax pair for the Adler (lattice Krichever-Novikov)
system. {\it Phys. Lett. A}, {\bf 297} (2002), no. 1-2, 49--58;
arXiv:nlin/0110027.

[4] A.I. Bobenko, Yu.B. Suris. Integrable systems on quad-graphs.
{\it Int. Math. Res. Notices}, {\bf 11} (2002), 573--611;
arXiv:nlin/0110004.

[5] V.E. Adler, A.I. Bobenko, Yu.B. Suris. Classification of
integrable equations on quad-graphs. The consistency approach. {\it
Comm. Math. Phys.}, {\bf 233} (2003), no. 3, 513--543;
arXiv:nlin/0202024.

[6] A.I. Bobenko, Yu.B. Suris. Discrete differential geometry.
Consistency as integra\-bility. arXiv: math/0504358.

[7] A.I. Bobenko, Yu.B. Suris. On organizing principles of discrete
differential geometry. Geometry of spheres. {\it Uspekhi Mat. Nauk},
{\bf 62}:1 (2007), 3--50; English translation in {\it Russian Math.
Surveys}. 2007. V. {\bf 62}, No. 1. P. 1--43; arXiv:math/0608291.

[8] A.P. Veselov. Integrable maps. {\it Uspekhi Mat. Nauk}, {\bf
46}:5 (1991), 3--45; English translation in {\it Russian Math.
Surveys}. 1991. V. {\bf 46}, No. 5. P. 1--51.

[9] V.E. Adler, A.I. Bobenko, Yu.B. Suris. Discrete nonlinear
hyperbolic equations. Classification of integrable cases. arXiv:
abs/0705.1663.

[10] S.P. Tsarev, Th. Wolf. Classification of three-dimensional
integrable scalar discrete equations. {\it Lett. Math. Phys}. {\bf
84} (2008), no. 1, 31--39. arXiv: abs/0706.2464.

[11] J. Hietarinta. A new two-dimensional lattice model that is
`consistent around a cube'. {\it J. Phys. A}, {\bf 37} (2004), no.
6, L67--L73; arXiv: nlin/0311034.

[12] V.E. Adler, A.P. Veselov. Cauchy problem for integrable
discrete equations on quad-graphs. Acta Appl. Math. 84 (2004), no.
2, 237--262; arXiv: math-ph/0211054.

[13] V.E. Adler, Yu.B. Suris. ${\rm Q}\sb 4$: integrable master
equation related to an elliptic curve. {\it Internat. Math. Research
Notices}, 2004, No. 47, 2523-2553; arXiv: nlin/0309030.

[14] I.M. Gelfand, M.M. Kapranov, A.V. Zelevinsky. Discriminants,
resultants, and multidimensional determinants. Mathematics: Theory
\& Applications. Birkhauser Boston, Inc., Boston, MA, 1994. x+523
pp.; Reprint of the 1994 edition. Modern Birkhauser Classics.
Birkhauser Boston, Inc., Boston, MA, 2008. x+523 pp.

\begin{flushleft}
{\bf O. I. Mokhov}\\
Centre for Nonlinear Studies,\\
L.D.Landau Institute for Theoretical Physics,\\
Russian Academy of Sciences,\\
Kosygina str., 2,\\
Moscow, Russia;\\
Department of Geometry and Topology,\\
Faculty of Mechanics and Mathematics,\\
M.V.Lomonosov Moscow State University,\\
Moscow, Russia\\
{\it E-mail\,}: mokhov@mi.ras.ru; mokhov@landau.ac.ru; mokhov@bk.ru\\
\end{flushleft}

\end{document}